\documentclass[aip,pop,amsmath,amssymb,reprint]{revtex4-1}

\def\eqalign#1{\null\,\vcenter{\openup\jot\m@th 
 \ialign{\strut\hfil$\displaystyle{##}$&$\displaystyle{{}##}$\hfil 
   \crcr#1\crcr}}\,} 

\usepackage{color}

\usepackage{epsfig}
\usepackage{natbib}
\usepackage{comment}
\usepackage{amsmath,amssymb,graphicx,url,gensymb,amsbsy}

\begin{document}

\title{Maximal energy extraction under discrete diffusive exchange}

\author{M. J. Hay}
\email[]{hay@princeton.edu}
\affiliation{Department of Astrophysical Sciences, Princeton University, Princeton, New Jersey 08544}

\author{J. Schiff}
\affiliation{Department of Mathematics, Bar-Ilan University, Ramat Gan, 52900 Israel}

\author{N. J. Fisch}
\affiliation{Department of Astrophysical Sciences, Princeton University, Princeton, New Jersey 08544}
\affiliation{Princeton Plasma Physics Laboratory, Princeton, New Jersey 08543}

\date{\today}

\begin{abstract}

Waves propagating through a bounded plasma can rearrange the densities of states in the six-dimensional velocity-configuration phase space.
Depending on the rearrangement, the wave energy can either increase or decrease, with the difference taken up by the total plasma energy.
In the case where the rearrangement is diffusive, only certain plasma states can be reached. 
It turns out that the set of reachable states through such diffusive rearrangements has been described in very different contexts.
Building upon those descriptions, and making use of the fact that the plasma energy is a linear functional of the state densities, the maximal extractable energy under diffusive rearrangement can then be addressed through linear programming. 
%
\end{abstract}


\maketitle

\section{Introduction}

Waves propagating through a bounded plasma can rearrange the densities of states in the six-dimensional velocity-configuration phase space.
When the rearrangement is such as to cause particles to diffuse from higher energy states to lower energy states, the waves extract energy from the plasma.
A particular case of this posing of the rearrangement problem is the case of alpha channeling, where the energy is deliberately extracted from the population of $\alpha$-particles that are produced in a fusion reactor.$\cite{fisch92}$ 
This energy is recovered as wave energy. 
In a reactor, this energy is more useful in the form of wave energy, which can be used to attain a hot-ion mode or to drive electrical current.

The rearrangements contemplated in plasma using waves are diffusive in nature, since the wave-particle mechanisms generally cannot maintain coherence, at least not for the leading way of using external rf sources to heat or drive current in plasma.
However, depending on the wave frequency and wavenumber, the diffusion occurs in paths that link energy to the spatial dimensions. 
In the case of tokamak reactors, where $\alpha$-particles are expected to be born at high energy in the plasma center, there is a natural energy inversion along the path that connects the dense phase space location at high energy in the center to the under-dense phase space location that is at low energy on the periphery. 
It is then only a matter of constructing the appropriate wave diffusion path to link these locations. 
How much energy can be released from $\alpha$-particles is a matter of considerable practical interest, since if appreciable energy could be released in this manner, there then might be the opportunity to diminish substantially the cost of electricity through tokamak fusion.

A theoretical issue of academic interest, however, is the precise maximum available energy under diffusive rearrangements of phase space density when an arbitrary number of diffusion paths can be constructed.\cite{fisch93} 
The energy extractable under diffusion was posed recently, in fact, as one of the interesting, outstanding problems in wave-particle physics in plasma.\cite{fisch13} 
This issue also motivated to some extent approaches to other bounds on energy exchange between light and plasma, such as the extent to which bounds could be placed on the absorption of laser light at an interface.\cite{levy14}
There are, to be sure, also other formulations of free energy in plasma under phase space density rearrangements. 
For example, respecting phase space conservation, the Gardner restacking\cite{gardner63,dodin05} represents a precisely definable free energy that can be readily calculated. This energy is an upper bound on that which can be extracted through diffusive processes. However, because the Gardner restacking is without the realistic limitation of the diffusion constraint, the free energy available under this formulation represents a rather rarefied theoretical construct, even further from practical considerations than the academic issue posed here.

The free energy under the constraint of diffusive rearrangements has one well-known textbook example, the famous so-called ``bump-on-tail" problem. 
The tail of the Maxwellian distribution is imagined to have a ``bump" in velocity space, so that the distribution is no longer monotonically decreasing in energy. 
In that case, waves can diffuse particles so as to smooth out the bump, releasing the kinetic energy, until a distribution monotonically decreasing in energy is reached. 
The maximum extractable energy is obvious, and can be constructed from geometrical considerations; for the 1-bump problem, it is just the energy change in flattening the bump. 
However, were there two bumps in the velocity space, then the optimal solution would no longer be obvious at all, since it is not clear which bump should be flattened first.

More generally, what is imagined here is that diffusion paths can be constructed that link any two phase space locations in the 6D velocity-configuration space, whether or not the locations are contiguous. \cite{fisch93}
The contiguous constraint can formally be realized, in any event, in the limit of vanishingly thin paths. 
Thus we imagine an ensemble of discrete phase space locations, each with an initial density, and each representing a certain energy. 
Then as population densities relax under diffusion, the total system energy relaxes as well. 
The free energy under the diffusion constraint is then defined as the maximum extractable energy, given the opportunity to diffuse particles between any two phase space locations, with any sequence of such two-location or what we might call two-state relaxations. 
It turns out, however, that while this pair-wise relaxation is a well-defined posing of the free energy, it has not been apparent at all how to calculate it efficiently when there are many states. 

The problem thus posed in plasmas can similarly be posed with respect to stimulated emission by a set of lasers.\cite{fisch93} 
Suppose, an atomic system with just three energy levels, the ground state at energy $\epsilon_1$, the first excited state at $\epsilon_2$,
and the second excited state at $\epsilon_3$, with initial population densities of, respectively, $n_0^1$, $n_0^2$, and $n_0^3$.  
The total energy can be put as $W=\vec{\epsilon} \cdot \vec{n}$, where $\vec{\epsilon}$ and $\vec{n}$ represent the energy levels and the population densities. 
Suppose
further the availability of three lasers with frequencies $\nu_{10}$, 
$\nu_{20}$, and $\nu_{21}$, that, respectively, can stimulate transitions
between the first level and the ground state, the second level and the ground
state, and the second level and the first level. Suppose that these lasers are incoherent, so what they can accomplish is to equalize the 
populations in any two levels. 
The maximum energy is extracted when the correct sequence of laser pulses is applied.

To make the issues here clear, consider the following example.\cite{fisch93} 
Suppose that the accessible states have energies with numerical values (0,1,4) and the initial state densities are (0,2/7,5/7), where the sum has been normalized to 1. 
Then the initial energy is $W_0$ = 22/7.
The energy at step $j$ can be put as $W_j$.  
The sequence of level-equalizing steps, $(\nu_{21},\nu_{20},\nu_{10})$, then gives
\begin{align*}
\bordermatrix{       &\epsilon_1=0 &\epsilon_2=1  &\epsilon_3=4\cr
{\rm initial}\qquad W_0=22/7 \qquad &0      &2/7       &5/7    \cr
{\rm step\ 1}\qquad W_1=5/2 \qquad &0      &1/2      &1/2   \cr
{\rm step\ 2}\qquad W_2=3/2\qquad &1/4     &1/2      &1/4   \cr
{\rm step\ 3}\qquad W_3=11/8 \qquad &3/8   &3/8     &1/4   \cr}
\end{align*} 
The energy extracted is thus $22/7 - 11/8=99/56$, or approximately $56\%$ of the initial plasma energy.  
One can show, for this set of energy levels and initial populations, that the sequence used is the optimal sequence for extracting energy, resulting in the maximum extractable energy. 
What is of interest, however, is how exactly this can be proved, how the maximum extraction can be calculated efficiently, and how the complexity of the problem increases with the number of states. 

It turns out that the answer to these questions lies in the mathematical developments in other fields (although these developments appear not to have received much attention).
Similar level-mixing operations have been considered in chemical reaction kinetics.\cite{horn64}
More directly of use here, Zylka identified the set of accessible states through level-mixing operations, called the $K$ set, using as an example the problem of attainable temperatures in heat reservoirs pairwise connected by heat pipes, with an arbitrary number of reservoirs.\cite{zylka85}  
Thon and Wallace, in the context of characterizations of altruism as a pairwise relaxation correction to economic inequality, derived further features of the $K$ set.\cite{thonwallace} 

Using these characterizations of the $K$ set, the sequence of operations to extract maximal energy can be found. Although $K$, the set of states that can be reached from these level-mixing operations, is not convex,\cite{zylka85} one can imagine covering the entirety of $K$ with a small convex polygon. This polygon (namely, the \emph{convex hull} of $K$, denoted $ch(K)$ here) is determined by taking all possible convex combinations of the points found within $K$. Equipped now with a linear objective function (the total energy of a state) and a convex feasible region (the unique covering polygon, $ch(K)$), we may apply the fundamental theorem of linear programming to locate the minimum energy state at a vertex or edge joining two or more vertices of the convex hull of $K$. Crucially, $K$ contains each of the vertices of its convex hull (by construction), and so the minimum energy state over the polygon \emph{covering} $K$ is identical to the minimum energy state found \emph{within} $K$.



Thus, using the results of Zylka and of Thon and Wallace, for the three-level system, we shall pose and answer the following five questions:

\begin{enumerate}
\item What sequence minimizes the energy?

\item What is the full set of sequences that must be considered before the optimal sequence can be found? 

\item  What are the full set of sequences that could possibly be a solution, for some values of the energy levels and the initial population densities? 

\item If it were possible to partially relax the distribution between two states, rather than fully relax it, would that ever be a useful step? 

\item Is it the case that it is ever useful to take a step that increases the energy rather than decreases it?

\end{enumerate}

The paper is organized as follows. In Sec.~II, we define the diffusion model. In Sec.~III, we reproduce results of the space of relaxation solutions, and show how this immediately answers the first three questions.  In Sec.~IV, we answer the fourth question, proving that partial relaxation is never a useful step, regardless of the number of states. In Sec.~V, we demonstrate out that strategies previously considered\cite{fisch93} can be put more precisely and further prove for any number of states that energy-increasing steps can never be part of the optimal sequence. In Sec.~VI, we offer further discussion of the implications for $N$ states and the degree of complexity of the problem.
In Sec.~VII, we summarize the main conclusions.

\section{Diffusion model}


Suppose a set of level energies $\epsilon=(\epsilon_1,\,\epsilon_2,\,\ldots,\,\epsilon_N)$ and a set of initial populations $n_0=(n^1_0,\,n^2_0,\,\ldots,\,n^N_0)$. 
Here we understand the symbols $\epsilon$ and $n_0$, without superscripts, to represent vectors; 
the initial level density of state $i$ is represented as $n_0^i$.
Consider a diffusion operation that equalizes the populations of a pair of levels $(i,\,j)$ such that
\begin{gather}
\left(n_0^i,\, n_0^j\right)\to\left(\frac{n_0^i+n_0^j}{2},\, \frac{n_0^i+n_0^j}{2}\right),
\label{transform}
\end{gather}
leaving all other level populations unchanged. 
We are interested here is the minimization of the system energy $W_d \doteq\epsilon\cdot n_f$
after repeated application of operations of this type on an initial state $n_0$ and reaching a state $n_f$. 

The diffusion operation \eqref{transform} can be represented by doubly stochastic matrices of the form $B_{ij}=\tfrac{1}{2}(I+Q_{ij})$, where $I$ is the $N\times N$ identity matrix and $Q_{ij}$ is the permutation matrix that exchanges the $i^{\rm th}$ and $j^{\rm th}$ level populations. 
Application of $B_{ij}$ 
equalizes the $i^{\rm th}$ and $j^{\rm th}$ populations. 
For example, \eqref{transform} could be styled $n_0\to n_0B_{ij}$.
The $B_{ij}$ are symmetric, idempotent, and do not generally commute. 
(Two $B_{ij}$ commute if neither of them operates on the same level.) 
They are a particular case of the $T$-transform: $T=(1-\alpha) I+\alpha Q,\,\alpha\in[0,1]$ and $Q$ is a permutation of the identity matrix which exchanges only two rows.\cite{marshallbook} 
Like all doubly stochastic matrices, $T$-transforms are measure-preserving: $\sum_{ab} T_{ab}n_b=\sum_b n_b=1$. 
It is important to distinguish between the cases of $0\leq\alpha\leq1/2$ and $1/2<\alpha\leq1$; the latter case corresponds to moving density from a less-populated level to a more-populated level. For now, we will consider only $\alpha=1/2$ transforms, i.e. the $B_{ij}$.

\section{The set of points $K$}

%
Zylka was the first to identify the set of accessible states through level-mixing operations: the $K$ set. \cite{zylka85} $K$ is determined by the provided set of level populations, represented as a length-$N$ vector $n_0$, as well as the allowed diffusion operations. Without loss of generality, $n_0$ may be ordered increasing, as we assume throughout this work. As described previously, the diffusive $B_{ij}$ operations equalize the populations of any pair of levels $i$ and $j$. Each element contained in $K$ is an $N$-tuple of level populations that can be reached by applying some sequence of the various $B_{ij}$ to the initial state $n_0$.

The linear function $W_d$ assumes its extremal values on the boundary of $ch(K)$.
The three-level system is conveniently depicted in $n_1$-$n_2$ space.
Due to normalization of the population vector, one coordinate is ignorable and a general state $n$ may be written as $n=(n_1,n_2,1-n_1-n_2)$;
if the total density is say 1, the the density of the third is simply $1- n_1-n_2$.
Following Zylka, we will demonstrate that $K$ is star-like.\cite{zylka85}   

To see this, first let us consider the example considered earlier, and previously,\cite{fisch93} namely the case of $N=3$ case with initial data $n_0=(0,2/7,5/7)$ and $\epsilon=(0,1,4)$.

\begin{figure}
\includegraphics[width=70mm]{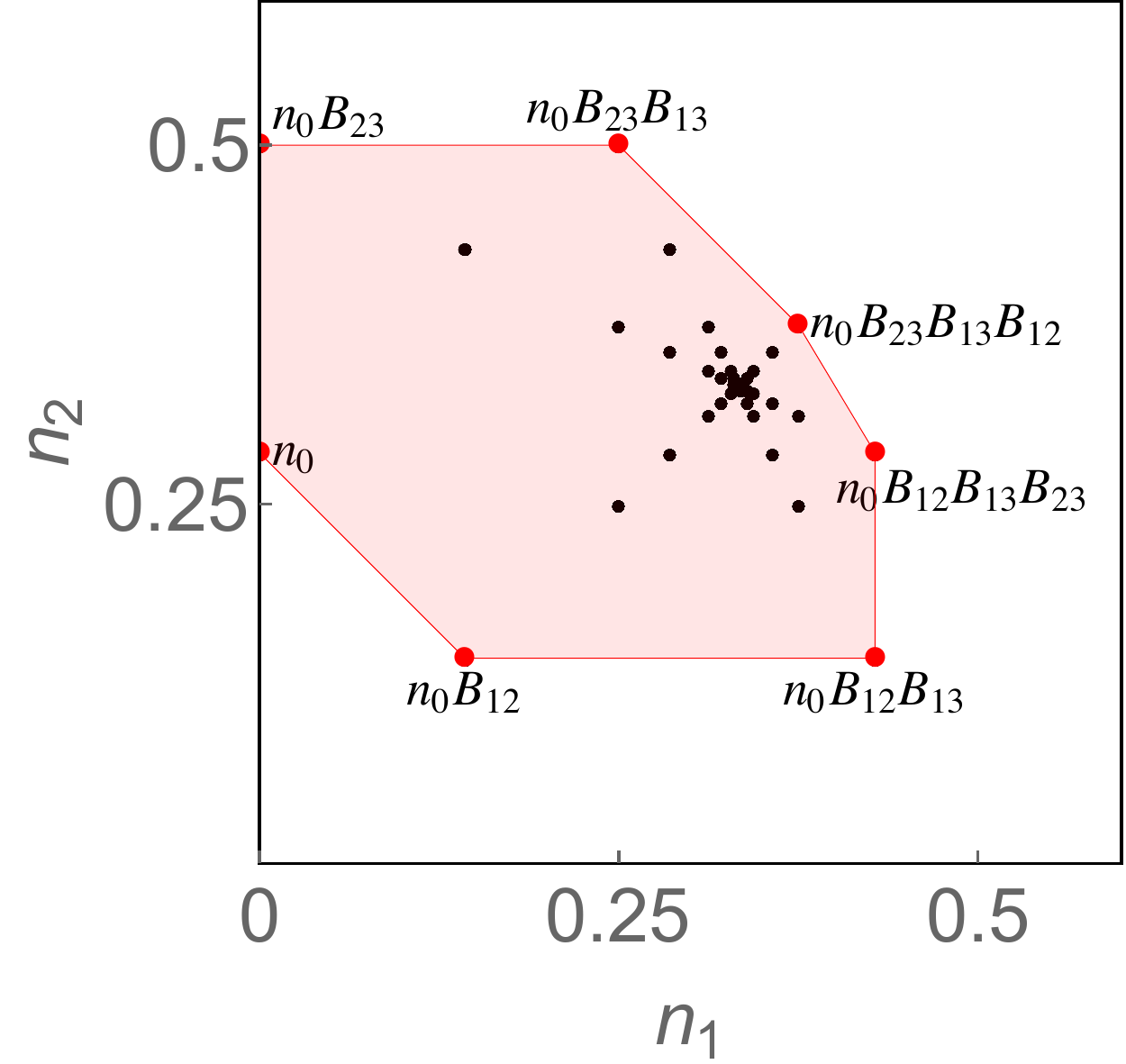}
\caption{The set $K$ for $n_0=(0,2/7,5/7)$, depicted in $n_1$-$n_2$ space. The convex hull $ch(K)$ is illustrated by the pink region. Extreme points of $ch(K)$ are indicated with red dots and labeled by the by the series of transformations $B_{ij}$ required to reach them. Interior points are labeled with small black dots. $K$ itself is the union of the red and black dots.}
\end{figure}

\begin{figure}
\includegraphics[width=70mm]{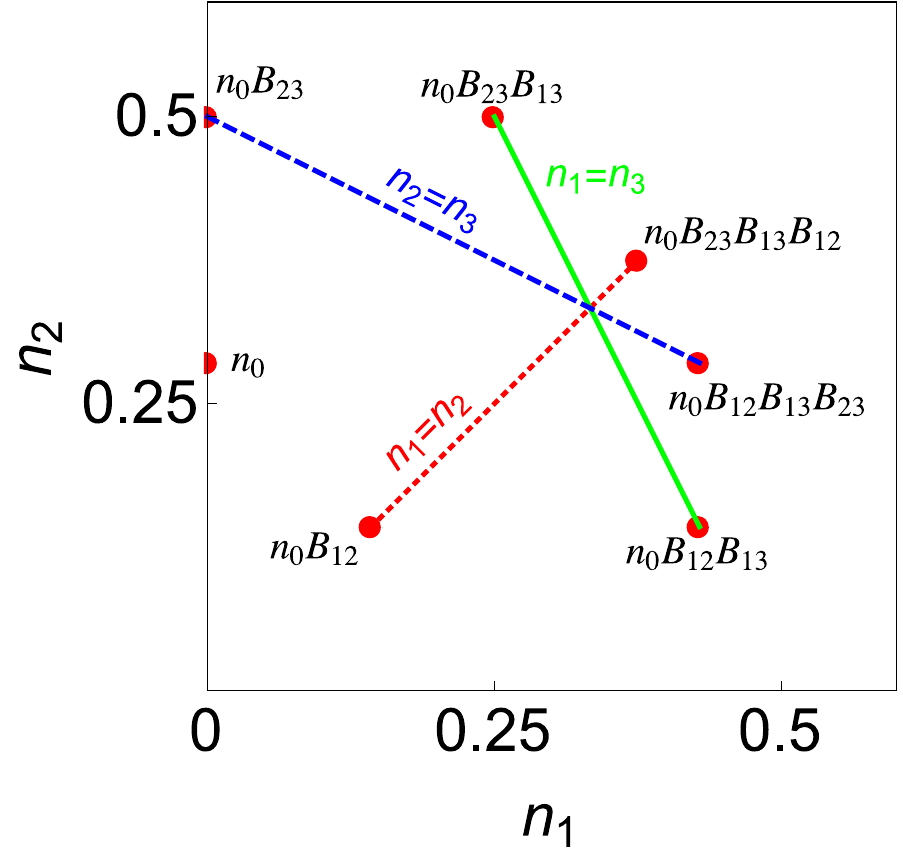}
\caption{The three lines covering $K$ are depicted for $n_0=(0,2/7,5/7)$, emphasizing the star-shaped nature of $K$.}
\end{figure}

One can generate the entire set $K$ by applying arbitrarily long sequences of the transforms $B_{ij}$ to the initial point $n_0$.
We denote the set of points generated by $k$ arbitrary $B_{ij}$ the $k^{\rm th}$ generation of $K$.
The first six generations of $K$ are plotted in Figure~1, overlaid with $ch(K)$. 
Except for the original point $n_0$, the entirety of $K$ lies along three distinct line segments joining pairs of extreme points, as depicted in Figure~2.
This arrangement is obvious if one considers that every state must have $n_i=n_j$, for some $i$ and $j$, if the last transform applied in the sequence arriving at that state is $B_{ij}$. 

The six extreme points (excepting $n_0$) can then be paired according to their final $B_{ij}$. 
For example, the extreme points $n_0 B_{12}B_{13}$ and $n_0 B_{23}B_{13}$ are so paired, and the line joining them satisfies $n_1=n_3=1-n_1-n_2$, or $n_2=1-2n_1$. 
It is now clear that $K$ must be star-shaped with respect to $e=(1/3,1/3,1/3)$ because the three lines joining pairs of extreme points satisfy $n_1=n_2=1-n_1-n_2$ there.\cite{zylka85} Likewise, $ch(K)$ for a three-level system is defined by at most seven extreme points (see Appendix~A).
Each generation approaches $e$ more closely, and no extreme points are generated after the third generation.
(Consider the effect of $B_{ij}$ on a state $n=(n_1,n_2,1-n_1-n_2)$. Then the Euclidean distance $s$ of $n$ from $e$ satisfies $s^2=(n-e)\cdot(n-e)$. Comparing the distances of $n$ and $n B_{ij}$ from $e$, one finds $(n-e)\cdot(n-e)-(nB_{ij}-e)\cdot(nB_{ij}-e)=\frac{1}{2}(n_i-n_j)^2\geq0$. Thus each $B_{ij}$ brings a state closer to $e$.)

\begin{figure}
\includegraphics[width=70mm]{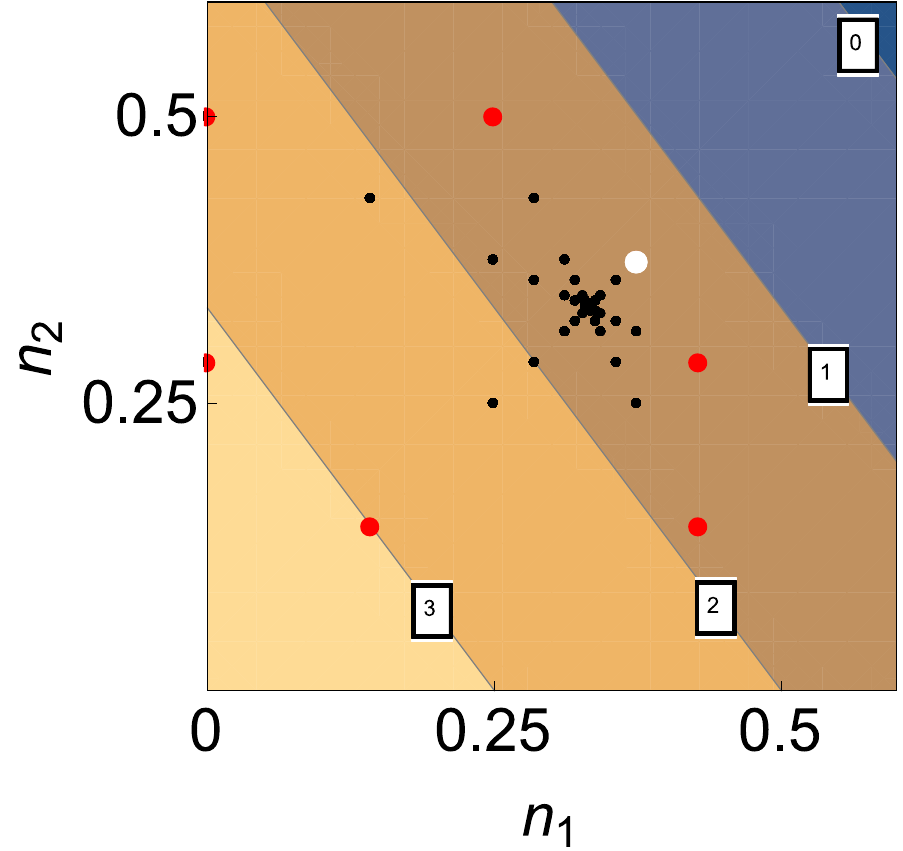}
\caption{$K$ superimposed on contours of $W_d=\epsilon\cdot n$ (labeled by energy). The minimum $W_d$ is found at the extreme point $n_0B_{23}B_{13}B_{12}=(3/8,3/8,1/4)$ (labeled by a red circle in Figs. 1 and 2) labeled here by a large white circle.}
\end{figure}
Note that the set $K$ is non-convex,\cite{zylka85} since not all of the points in $ch(K)$ can be reached by sequences of $B_{ij}$ applied to $n_0$.
Per the fundamental theorem of linear programming, the minimum $W_d$ over the convex feasible region $ch(K)$ is found either at a single extreme point of $ch(K)$ or along an edge of $ch(K)$ joining two or more extreme points of $ch(K)$. (In the latter case, the minimum $W_d$ multiplicity is equal to the number of collinear extreme points.)  However, the extreme points of $ch(K)$ are all contained in $K$. Thus the linearity of the objective function $W_d=\epsilon\cdot n$ ensures that it has the same minimum over $K$ and $ch(K)$.


Having identified the set of extreme points of $ch(K)$, $\mathcal{E}(ch(K))$, in Fig.~1, the minimum $W_d$ may be computed as ${\rm min}\{\epsilon\cdot \mathcal{E}(ch(K))\}$. Fig.~3 overlays contours of the state energy $\epsilon\cdot n$ on $K$, thereby illustrating the scheme and identifying the minimum energy state. In this case, the optimum level populations are $n_f=n_0B_{23}B_{13}B_{12}=(3/8,3/8,1/4)$, yielding a minimum $W_d=11/8$, as previously calculated by exhaustive search.

The sequence which minimizes the plasma energy can always be identified with this algorithm.
Crucially, because we have not assumed any ordering of the level energies $\epsilon$, the slope of the state energy contours is arbitrary in the general case (cf. Fig.~3). Therefore, each of the seven extreme points identified is a possible solution to the energy minimization problem.

\section{Partial relaxation}

The discussion so far has assumed complete pairwise equalization of levels at every step, corresponding to $B_{ij}$ transforms with $\alpha=1/2$.
From a physical standpoint, densities could also be relaxed through a diffusion process that only partially equalizes population levels, corresponding to the case of $T$-transforms with $0<\alpha<1/2$. 

The $\alpha=1/2$ transforms $B_{ij}$ can be reached by repeatedly applying a particular $T$-transform with $0<\alpha<1$ formed from the same $Q_{ij}$, since for all $0<\alpha<1$, $\lim_{n\to\infty}T_{ij}^n=\tfrac{1}{2}(I+Q_{ij})=B_{ij}$. 
Thus, the $\alpha=1/2$ state space is in the closure of the state space for any fixed $0<\alpha<1$. 
At the same time, the $0<\alpha<1/2$ state space is contained entirely in the convex hull of the $\alpha=1/2$ state space.
To see this, simply observe that for $0<\alpha<1/2$,
\begin{gather}
T_{ij}=(1-\alpha)I+\alpha Q_{ij}=(1-2\alpha)I+2 \alpha B_{ij},
\end{gather}
where $0<2\alpha<1$.
Thus any $0<\alpha<1/2$ transform (on a state in the $\alpha=1/2$ state space) results in a state that is a convex combination of the initital state and the state resulting from the corresponding $\alpha=1/2$ transform, i.e. an interior point of $ch(K)$. See Fig.~4 for a depiction of these possibilities. For any $T_{ij}$ with $0<\alpha<1/2$, the new state is constrained to lie on the dashed line joining $p$ and the corresponding $pB_{ij}$.

\begin{figure}
\includegraphics[width=70mm]{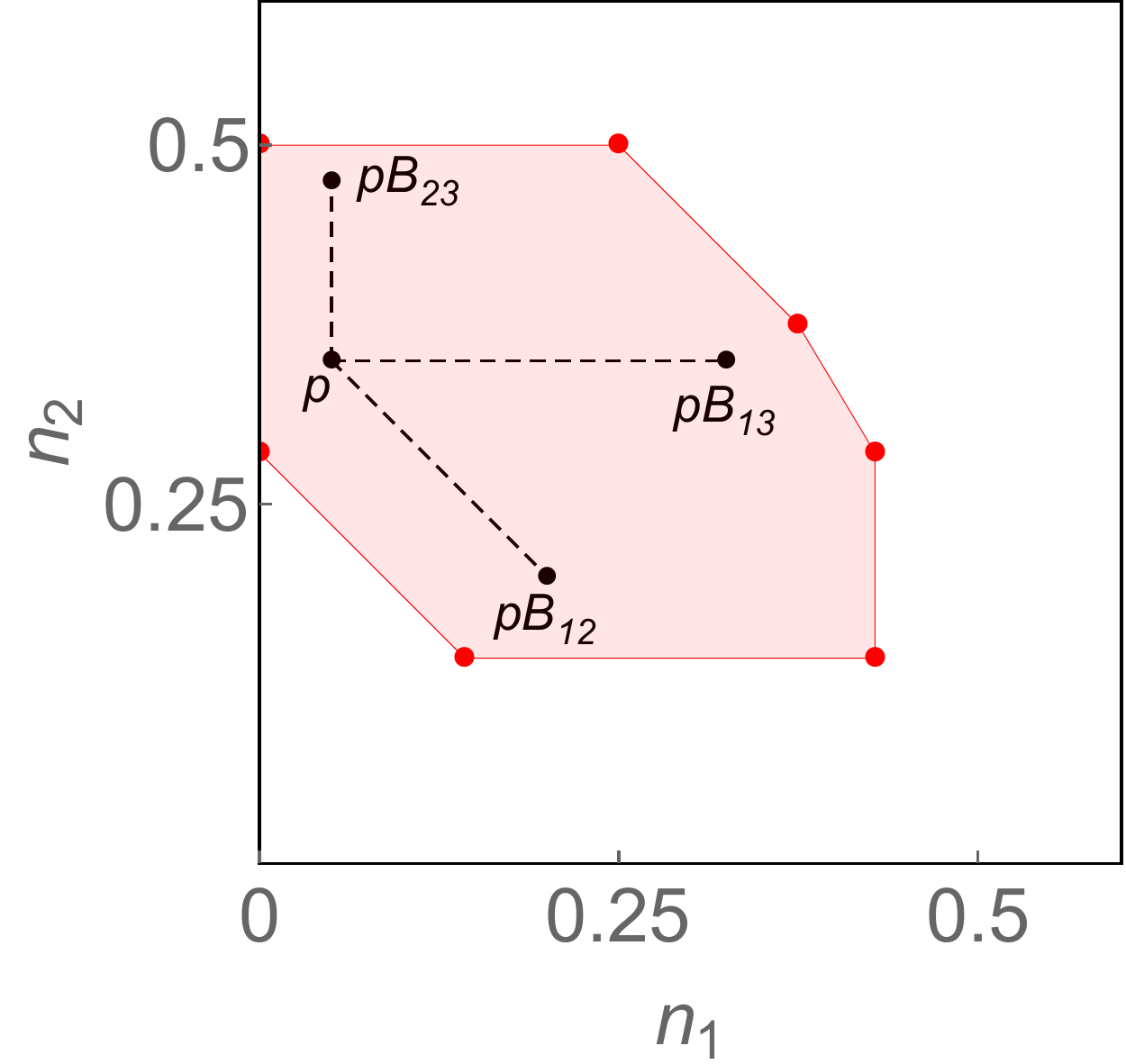}
\caption{Result of partial relaxations ($T_{ij}$ with $0<\alpha<1/2$) applied to a point $p$ in the convex hull of $K$ for $n_0=(0,2/7,5/7)$, as before. Partial relaxations result in interior points constrained to lie along the dashed lines joining $p$ to each $pB_{ij}$.  }
\end{figure}

It follows that it is sufficient to consider only $\alpha=1/2$ transforms to identify the extreme points of the $0<\alpha\leq1/2$ state space and accordingly solve the energy minimization problem.
\section{Strategies for Finding the Optimum Sequence} 


Two strategies for determining the sequence of transforms $B_{ij}$ required to attain the minimum-$W_d$ state were conjectured.\cite{fisch93} 
Reproduced here, these strategies are:
\begin{enumerate}
\item Diffusion of particles first between similar population levels, all other things being equal, eventually releases more energy.
\item Depleting of particles the higher energy level first, all other things being equal, eventually releases more energy.
\end{enumerate}

These strategies were only surmised based on trial-and-error experience. 
However, it turns out that they are intimately related to the more precisely put {\it Proposition 2}, proposed and proven by Thon and Wallace.\cite{thonwallace} It is worthwhile here to make use of the \emph{permutation} of the set of initial populations, as introduced in Ref.~\citenum{thonwallace}. Given $n_0=(n_1,n_2,n_3)$, the starting permutation is $\{1,2,3\}$ if $n_1<n_2<n_3$, as assumed throughout this work. Application of $B_{ij}$ exchanges the level numberings, e.g. if $n_0$ has permutation $\{1,2,3\}$, $n_0B_{12}$ has permutation $\{2,1,3\}$. For a more detailed exposition of these permutations and emergent combinatorial techniques, please refer to Appendix~B.

Put briefly, {\it Prop. 2} states that extreme points can be obtained by equalizing only \emph{adjacent} level pairs. That is, any sequence of $B_{ij}$ resulting in an extreme point will effect only adjacent transpositions in the population permutation. More concretely, $n_0B_{12}$ and $n_0B_{23}$ are both extreme points, but $n_0B_{13}$, which averages the nonadjacent first and third levels, is not. However, $n_0B_{12}B_{13}$ is an extreme point because the later application of $B_{13}$ equalizes two adjacent levels, viz. $n_0\sim\{1,2,3\}\to n_0B_{12}\sim\{2,1,3\}\to n_0B_{12}B_{13}\sim\{2,3,1\}$.

Because the initial populations are assumed ordered increasing and the global minimum energy state will always be located at an extreme point, any correct first step must be consistent with Strategy 1: the $B_{ij}$ chosen must mix two adjacent levels.

It is clear that there are `dead ends' among the possible $B_{ij}$ sequences, where a state is reached with level populations decreasing with level energy, such that no more energy can be extracted. Any such {\it stopping} state has the level population permutation which is the reverse of the energy level permutation. For example, given $\epsilon=(\epsilon_1,\epsilon_2,\epsilon_3)$ with $\epsilon_2<\epsilon_1<\epsilon_3\sim\{2,1,3\}$, the {\it stopping permutation} is $\{3,1,2\}$, such that $n_3\leq n_1\leq n_2$. $n\sim\{3,1,2\}$ can be reached from $n_0\sim\{1,2,3\}$ by the sequence $n_0B_{23}B_{13}$. 

The three-level system has two
extreme points with permutation $\{3, 2, 1\}$, the stopping
permutation if the level densities and energies are both
ordered increasing. In one case, the two lower-energy levels are first diffused; in the other, the two higher-energy
levels are first diffused. The state energies of the extreme points cannot be ordered using a reduced set of variables (i.e. only some of the initial populations or energy levels), so the correct first step can only be determined in retrospect after a full calculation. The multiplicity of extreme points with a possible stopping permutation indicates that Strategy 2 is not generally applicable.

%

A further question of interest is the usefulness of an `annealing' strategy whereby a diffusion operation heats the system, resulting in a state with extra energy. 
Subsequent $B_{ij}$ would lower the system energy to its minimum value, presumably smaller than the minimum value possible without annealing. However, such a strategy cannot obtain a global minimum $W_d$. Note that any step which heats the system results in an inversion of the population permutation with respect to the correct stopping permutation. 
In order to obtain the stopping permutation, a subsequent diffusion operation is required on the two levels involved the heating step.

As an example, consider a case mentioned previously with level energy permutation $\epsilon\sim\{2,1,3\}$, initial population permutation $n_0\sim\{1,2,3\}$, and stopping permutation $n\sim\{3,1,2\}$. Note that the initial permutation contains two inversions with respect to the stopping permutation. Applying $B_{12}$ to the initial state results in the system absorbing energy,
\begin{gather}
\epsilon\cdot n_0 B_{12}-\epsilon\cdot n_0=\frac{1}{2}(n_2-n_1)(\epsilon_1-\epsilon_2).
\label{de}
\end{gather}
By construction, each difference on the right hand side of Eq.~\eqref{de} is positive and the system absorbs energy. Now consider the number of inversions: $n_0B_{12}$ results in the level population permutation $n\sim\{2,1,3\}$, containing a total of three inversions with respect to the stopping permutation $\{3,1,2\}$. The extra inversion changes the parity of the permutation and requires at least one additional diffusion operation to correct. In particular, the same diffusion operation $B_{12}$ must be repeated at some point in the relaxation process. Per \emph{Prop. 3} of Ref.~\onlinecite{thonwallace}, such a $B_{ij}$ sequence does not generate an extreme point of $ch(K)$, even if both instances of the repeated $B_{ij}$ correspond to adjacent transpositions. However, there exists at least one extreme point with the stopping permutation. Because $K$ is star-shaped, any annealing strategy results in a non-optimal interior point (a convex combination of the uniform distribution $e$ and an extreme point with the stopping permutation). Because the parity and number of inversions characterize any finite-length permutation, this conclusion holds in the general $N$-level case.

\section{Complexity} 

Given length-$N$ initial data, there are at most ${N \choose 2}$ possible states after one transformation has been applied (fewer in case of degeneracy in the initial level populations). The second generation contributes up to $N(N-1)(N-2)(N+5)/8$ unique states.\cite{oeis_triangle} Although the number of possible unique states grows rapidly with the passing generations, these later states spiral quickly toward the uniform distribution $e=(1/N,1/N,\ldots,1/N)$. In fact, we can safely restrict our attention to the first ${N \choose 2}$ generations, a nevertheless enormous set for large $N$. 

Thon and Wallace\cite{thonwallace} presented an algorithm for generating the extreme points of $ch(K)$. In fact, the algorithm identifies the reduced (minimum length) sequences of adjacent transpositions (i.e. permutations exchanging only neighboring elements in the set) leading to each permutation in the $N^{\rm th}$ symmetric group $\mathcal{S}_N$. For example, if $N=3$, the group $\mathcal{S}_3$ contains $N!=6$ possible permutations. The permutation $\{3,2,1\}$ can be reached from the initial $X=\{1,2,3\}$ by two distinct sequences of adjacent transpositions: $(1,2)(2,3)(1,2)$ and $(2,3)(1,2)(2,3)$. The other five possible permutations of $X$ have unique reduced decompositions in adjacent transpositions; the total number of such decompositions for all of the permutations in $\mathcal{S}_3$ is therefore seven.\cite{oeis_reduced} As noted, each possible permutation in $\mathcal{S}_N$ is a stopping state for an appropriate permutation of $\epsilon$.

Any sequence of diffusion operations with a repeated $B_{ij}$ is not minimal and results in a permutation accessible with ${N \choose 2}$ or fewer $B_{ij}$. This follows because the reverse permutation, the longest permutation when expressed in adjacent transpositions, requires precisely ${N\choose2}$ transpositions (corresponding to the number of inversions in the final permutation). For example, $nB_{23}B_{13}B_{12}B_{23}$ has the permutation $\{2,3,1\}$, which could also be reached with $nB_{12}B_{13}$.

In the $N=3$ case, the minimization of $W_d$ is straightforward. There are ${3\choose 2}=3$ unique $B_{ij}$: $B_{12},\,B_{13},\,{\rm and}\,B_{23}$. Without loss of generality, the initial data $n_0$ is increasing, so the initial permutation is $\{1,2,3\}$. The set $K$ is covered by $ch(K)$, which has seven extreme points: $n_0$, $n_0B_{12}$, $n_0B_{23}$, $n_0B_{12}B_{13}$, $n_0B_{23}B_{13}$, $n_0B_{12}B_{13}B_{23}$, and $n_0B_{23}B_{13}B_{12}$.

There are $\epsilon$ for which each of the seven extreme points can serve as the optimum system configuration. As noted, $\epsilon\sim\{1,2,3\}$ is a special case in which the decomposition has maximum length ${3\choose 2}=3$; there are two unique decompositions of the stopping permutation $\{3,2,1\}$ in adjacent transpositions. Equivalently, there are two extreme points with the stopping permutation: the minimum system energy is then $W_d={\rm min}\{\epsilon\cdot n_0B_{23}B_{13}B_{12},\epsilon\cdot n_0B_{12}B_{13}B_{23}\}$. This is the most general scenario for the free energy optimization problem, in which multiple extreme points with the correct stopping permutation must be compared.


The problem has been reduced to evaluating the function $W_d=\epsilon\cdot n$ at a finite number of known extreme points. 
Unfortunately, the upper bound on the number of extreme points with stopping permutations is $\mathcal{O}(N^{N^2})$.\cite{oeis_permutations,stanley84} 
Because there is no \emph{a priori} means of ordering these points in energy, the exponential depth of the state tree is intrinsic to the optimization problem.\cite{yannakakis} 
Thus calculating the system energy accessible with discrete diffusive exchanges is NP-hard. 

\section{Conclusions} 

By utilizing the (mostly ignored) literature on pairwise relaxation transformations, it is a relatively simple matter to apply the techniques of convex optimization and combinatorics to answer the questions posed. In particular, we find that for the three-level case, only seven sequences need be considered. One further conclusion that can be drawn is that each of the sequences that must be considered will definitely be the solution at least for one energy vector. In general, a large but finite number of sequences must be checked to solve the energy extraction problem. Moreover, partial relaxation or annealing strategies are never useful. 

The exposition here has also laid the foundation for analyzing the extremal properties of more complicated systems. One might make use of the results about $K$ to extremize nonlinear objective functions, as would be necessary in e.g. systems with electrostatic self-energy. Alternately, formulating the optimization problem over a system of countable particles (as in integer programming) could shed light on the manipulation of degenerate matter. These systems and others like them could reveal the physical significance of the non-convexity of the state space.


Although we have shown that identifying the maximal energy extraction solution is impractical in most cases of interest, the results obtained along the way nonetheless provide useful constraints on diffusive schemes that should narrow the search for efficient strategies.



 \begin{acknowledgments}
The authors are very grateful for discussions with L. Shi and S. Davidovits. Particular thanks are due to Professor D. Tannor for critical discussions at an early stage of this work.  Work supported by DOE Contract No. DE-AC02-09CH11466 and DOE NNSA SSAA Grant No. DE274-FG52-08NA28553. One of us (NJF) acknowledges the hospitality of the Weizmann Institute of Science, where he held a Weston Visiting Professorship during the time over which this work was initiated.
 \end{acknowledgments}

\bibliography{bibliography}

\appendix

\section{Extreme point geometry}
\begin{figure}
\includegraphics[width=70mm]{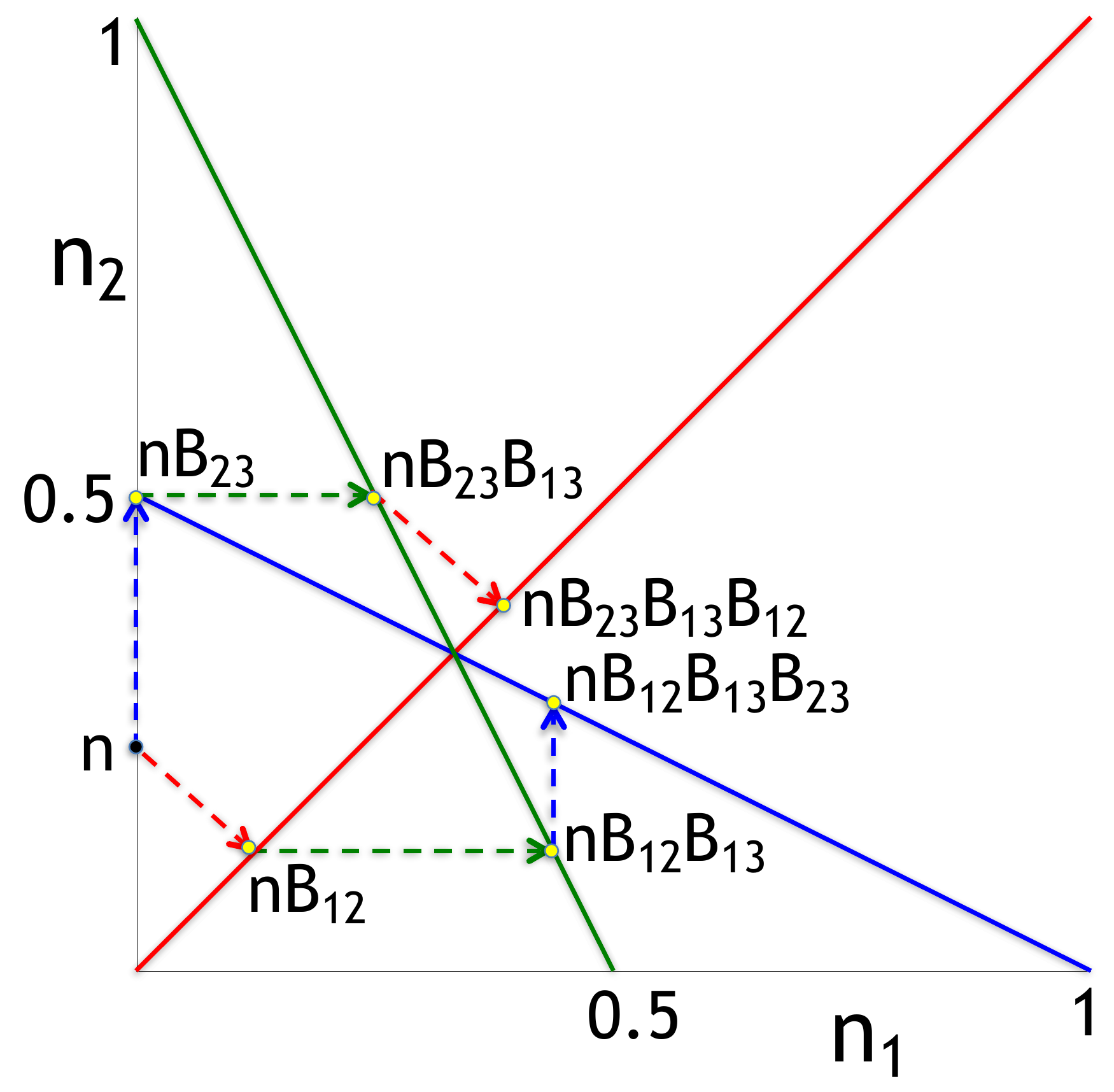}
\caption{Geometric construction of $K$. Extreme points of $ch(K)$ (yellow dots) depicted with lines $n_i=n_j$ (solid) and lines joining extreme points (dashed).\label{lines}}
\end{figure}
Thon and Wallace proved\cite{thonwallace} that their algorithm generates the full set of extreme points for an arbitrary number of energy levels $N$. However, the $N=3$ case admits a straightforward geometric proof. 
All states resulting from the application of a $B_{ij}$ lie along the line $n_i=n_j$. 
This constrains the possible paths through state space. In particular, the slope of the line joining any two points $p_1$ and $p_2$ such that $p_2=p_1B_{ij}$ is $-1$ if $(i,j)=(1,2)$, $0$ if $(i,j)=(1,3)$, or $\infty$ if $(i,j)=(2,3)$.

Fig.~\ref{lines} plots the extreme points and the three lines covering $K$ as well as these constrained trajectories. 
One extreme point is given ($n$, the initial data), and the other six are generated from choosing one of two routes through state space. 
Note that extreme points are only obtained by moving directly to a line $n_i=n_j$, without crossing any other such line. 
The lines are color-coded. 
Thus, for example, the partial relaxation along a red dotted line terminates on the red solid line; the red solid indicates complete relaxation of levels 
$n_1$ and $n_2$, while all the states traversed along the dotted line are reachable.
It can be seen from this geometrical representation that no extreme points are obtained by revisiting the same line: each diffusive transformation $B_{ij}$ reduces the distance of the state from the uniform distribution $e$. 
Thus, a state that revisits a line would necessarily be interior to the first state reached on that line.

\section{Combinatorial methods}
Thon and Wallace\cite{thonwallace} simplified the structures of many proofs by making use of the combinatorial features of the Dalton transfer problem. In particular, they introduced the permutation of the individuals whose incomes were to be redistributed. In their notation, the initial permutation is denoted $v=(v(1),v(2),\ldots,v(N))$, where $v(i)=k$ is interpreted to mean that the individual numbered $k$ is the $i^{\rm th}$ poorest. This representation is equivalent to the common {\it two-line notation}:
\[  \left( \begin{array}{cccc}
1 & 2 & \cdots & N \\
v(1) & v(2) & \cdots & v(N) \\
\end{array} \right),\]
where the first row lists indices of the permutation and the second row identifies the individuals. Note that the ordering of the columns is immaterial. The first column could be read as ``individual $v(1)$ is the poorest, listed first."

In their problem, as in ours, it is assumed that the individuals (levels) are ordered initially from poorest (least populated) to richest (most populated), yielding an initial permutation $v(i)=i$. Because the diffusion operations treated in this work transfer level densities but leave the level energies unchanged, we lose no generality by specializing to such a permutation, which we denote $\{1,2,\ldots,N\}$, reflecting $n_0^1\leq n_0^2\leq\cdots\leq n_0^N$. (This is the {\it ordered arrangement} or {\it one-line} representation of the level density permutation.) In concordance with Ref.~\citenum{thonwallace}, we describe the levels nearest in population as {\it neighbors}, e.g. $n_2$ is neighbors with $n_1$ and $n_3$ only. It turns out that the extreme points of $ch(K)$ can be reached only by sequences of $B_{ij}$ averaging such neighboring levels ({\it Prop. 2} of Ref.~\citenum{thonwallace}).

Hoping for greater simplicity, we discuss separately the permutation of the level energies, denoted $\{w(1),w(2),\ldots,w(N)\}$, with $w(i)=k$ meaning the level with the $k^{\rm th}$ lowest energy has initially the $i^{\rm th}$ smallest population. For example, if there is a complete population inversion, the highest-energy levels are initially most populated and $\epsilon\sim\{1,2,\ldots,N\}$. If instead $N=3$ and the second-highest energy level is most populated, followed in turn by the lowest and highest energy levels, one has $\epsilon\sim\{3,1,2\}$, or, in two-line notation:
\[  
\left( \begin{array}{ccc}
3 & 2 & 1 \\
2 & 1 & 3 \\
\end{array} \right)
=\left( \begin{array}{ccc}
1 & 2 & 3 \\
3 & 1 & 2 \\
\end{array} \right).\]
(Recall our assumption $i=v(i)$, so that the first row of the two-line representation of the energy permutation identifies the states ordered by initial level populations. Thus the columns of the two-line representation could be parsed ``greatest density in second-highest energy level, second-highest density in the lowest energy level, and least density in highest-energy level.") 

Apart from equalizing the populations of levels $i$ and $j$, one can consider the effect of the $B_{ij}$ on these permutations. As mentioned previously, the $B_{ij}$ have no effect on the permutation of level energies, which remains fixed throughout the problem. However, each $B_{ij}$ changes the permutation of level populations, with the result that the numberings are exchanged, so that e.g. $n_0B_{13}$ has permutation $\{3,2,1\}$. Considering the problem in $n_1$-$n_2$ space, one realizes that the effect of a $B_{ij}$ is to move the system state to another of the $N!=6$ cells in $ch(K)$, each corresponding to a specific permutation of the level populations (cf. Fig.~1 of Ref.~\citenum{thonwallace}). The cells are separated by rays from $e$ to six of extreme points in $ch(K)$ ($n_0$ excluded).

One of the key insights of this approach is the connection between population inversions and inversions in the permutation of level densities vis-\'a-vis the energy permutation. Using only the $B_{ij}$, it is possible to reorder the entire set of level populations, so that the final distribution is decreasing with energy (we term the permutations corresponding to such distributions {\it stopping} permutations).

In fact, there is a unique stopping permutation in a given problem: the {\it reverse} of the energy permutation, i.e. $\{w(N),w(N-1),\ldots,w(1)\}$, obtained from the composition of $\mathbf{w}$ with the order-reversing permutation. Let $\mathbf{w}^r$ denote the reverse of the permutation $\mathbf{w}$. In two-line notation, one has
\[ \mathbf{w}^r = \left( \begin{array}{cccc}
1 & 2 & \cdots & N \\
w(1) & w(2) & \cdots & w(N) \\
\end{array} \right)
\left( \begin{array}{cccc}
1 & 2 & \cdots & N \\
N & N-1 & \cdots & 1 \\
\end{array} \right)\]

\[  =\left( \begin{array}{cccc}
N & N-1 & \cdots & 1 \\
w(N) & w(N-1) & \cdots & w(1) \\
\end{array} \right)
\left( \begin{array}{cccc}
1 & 2 & \cdots & N \\
N & N-1 & \cdots & 1 \\
\end{array} \right)\]
\[  =\left( \begin{array}{cccc}
1 & 2 & \cdots & N \\
w(N) & w(N-1) & \cdots & w(1) \\
\end{array} \right).\\
\]
Note that the reverse operation is an involution, such that $(\mathbf{w}^r)^r=\mathbf{w}$. For example, $\{1,2,3\}$ and $\{3,2,1\}$ are reverse permutations. In our previous example, the level energy permutation was $\{3,1,2\}$ such that $\epsilon_3\leq \epsilon_1 \leq \epsilon_2$ for $\epsilon=(\epsilon_1,\epsilon_2,\epsilon_3)$ and $n_0=(n_0^1,n_0^2,n_0^3)$. The reverse of the $\epsilon$ permutation is $\{2,1,3\}$. Thus any stopping state has $n_2\leq n_1 \leq n_3$, such that the final population in the highest-energy level (`level 2,' when ordered by initial population) is smallest, and so on.

By identifying the level energies' permutation and its reverse, the search for the correct sequence of $B_{ij}$ can be greatly narrowed because only extreme points with the stopping permutation need be considered.


\end{document}